\documentclass[11pt]{article}

\usepackage{cite}
\usepackage{graphicx}
\usepackage{amsmath,amssymb}
\usepackage[body={6in,8.5in}]{geometry}
\usepackage{color}

\usepackage{subfigure}
\usepackage{url}
\usepackage{authblk}

\title{Arctic melt ponds and bifurcations in the climate system}

\author[1]{Ivan Sudakov \thanks{sudakov@math.utah.edu}}
\author[2,3]{Sergey A. Vakulenko}
\author[1]{Kenneth M. Golden}
\affil[1]{Department of Mathematics, University of Utah}
\affil[2]{Institute of Problems in Mechanical Engineering, Russian Academy of Sciences}
\affil[3]{University ITMO}
\bigskip
\date{}
 
\begin{document}
\maketitle
\begin{abstract}
Understanding how sea ice melts is critical to climate projections.
In the Arctic, melt ponds that develop on the surface of sea ice 
floes during the late spring and summer largely
determine their albedo -- a key parameter in climate modeling.
Here we explore the possibility of a conceptual sea ice climate model 
passing through a bifurcation point -- an irreversible critical 
threshold as the system warms, by incorporating geometric information
about melt pond evolution.
This study is based on a  bifurcation analysis of the energy balance climate model 
with ice - albedo feedback as the key mechanism 
driving the system to  bifurcation points. 

\end{abstract}

\section{Introduction}

Sea ice is not only a sensitive, leading indicator of climate change, 
it is a key player in Earth's climate system. It also serves as a primary 
habitat for algal and bacterial communities which sustain life in the polar oceans. 
Perhaps the most visible, large scale change on Earth's surface in recent 
decades has been the precipitous decline of summer Arctic sea ice. 
With this significant loss of a white reflecting surface covering the Arctic Ocean, 
its albedo or reflectance decreases, and solar radiation is absorbed by the 
ocean rather than being reflected. This heats the upper ocean, melting 
even more ice, and so on, which is known as "ice-albedo feedback".

While global climate models predict a general decline in 
Arctic sea ice over the 21$^{\text{st}}$ century, 
the observed losses have significantly outpaced 
projections \cite{Stroeve:GRL-2007,Ped09}.
Improving our predictive capability for the fate of Earth's sea ice
cover and its ecosystems depends on
a better understanding of important processes and feedback mechanisms.
For example, during the melt season the Arctic sea ice cover
becomes a complex, evolving mosaic of ice,
melt ponds, and open water.
The albedo of sea ice floes is determined by melt pond configurations
\cite{PerN,Felt2010,Pol12}.
As ponds develop,
{\it ice-albedo feedback}
enhances the melting process.
Understanding such mechanisms and their impact
on sea ice evolution and its role 
in the climate system is critical to advancing how sea ice
is treated in climate models and improving projections.

Conceptual, or {\it low order} 
climate models often introduce feedback  through
empirical parameterization, for example, 
taking into account a simple relation between 
temperature and area of ice covered surface.  
There is a wide range of such works, including
 \cite {Nor75,Fraed79,Cur95,Eis09}. Usually, ice-albedo 
feedback was simply associated with a decrease in 
ice covered area and a corresponding increase in the 
surface temperature, further decreasing the ice covered area. 
Given the key role that melt pond formation and evolution plays
in sea ice albedo, we note here an apparent lack of incorporation
of such features into conceptual models of ice-albedo feedback.
Here we note that it is important to explore how melt pond geometry and
thermodynamics affect conceptual climate models, 
and ice-albedo feedback in particular.

While melt ponds 
form a key component of the Arctic marine environment, 
comprehensive observations or theories of their formation, 
coverage, and evolution remain relatively sparse. 
Available observations of melt ponds show that their 
areal coverage is highly variable, particularly for 
first year ice early in the melt season, with rates of 
change as high as 35 percent 
per day \cite{Pol12}. 
Such variability, as well as the influence of many 
competing factors controlling melt pond and ice floe 
evolution, make  realistic treatments of ice-albedo feeedback 
in climate models quite challenging \cite{Pol12}. 
Small and medium scale models of melt ponds which 
include some of these mechanisms have been 
developed \cite{Sky07,Felt2010}, and melt pond 
parameterizations are being incorporated into global 
climate models \cite{Ped09}. 

Moreover, recently 
it has been found \cite{Hoh12} that melt pond geometry has a 
complex fractal
structure, and that the fractal dimension exhibits a transition
from 1 to about 2 around a critical length scale of 100 m$^2$ in area.
This behavior should be taken 
into account in investigating sea ice-albedo feedback.  

Given the complex, highly nonlinear character of the underlying 
differential equations describing climate, 
it is natural to ask whether the decline of summer Arctic sea ice has 
passed through a so-called {\it tipping point}, or irreversible 
critical threshold as the system progresses toward ice-free summers
\cite{Eis09,Abb11}.
A key mechanism potentially driving the system to "tip" is 
ice-albedo feedback. The main aim of this work is to investigate
such a tipping point for a simplified model of sea ice 
and the climate system which takes into some account the 
evolution of melt pond geometry and its effect on sea ice albedo.

The surface 
of an ice floe is viewed here as a two phase composite 
of dark melt ponds and white snow or ice. The onset of 
ponding and the rapid increase in coverage beyond the 
initial threshold is similar to critical phenomena 
in the theory of phase transitions. Here we ask if 
the evolution of melt pond geometry $-$ and sea ice albedo $-$ 
exhibit universal 
characteristics which do not necessarily depend on the details of the driving mechanisms in numerical melt pond models. 
Fundamentally, the melting of Arctic sea ice is a phase 
transition phenomenon, where a solid turns to liquid, 
albeit on large regional scales and over a period of time 
which depends on environmental forcing and other factors. 
We thus look for features which are 
mathematically analogous to related phenomena in the theories 
of phase transitions and composite materials.

Basing our approach on the standard nonlinear phase transition model in the 2D case \cite{caginalp89}, 
we propose an expression for the rate of change of the melt pond size. It can be extended to 
the 3D case taking into account the vertical transfer of water to the ocean through ice due to the different physical processes.  
After that, we introduce the expression for albedo of the 
ice-covered surface and investigate through the melt pond size how the unexpected 
fractal geometry of  
melt ponds \cite{Hoh12} can influence the formula for albedo of the ice covered surface. 

As the next step, we consider a standard conceptual climate model--
an ordinary differential equation (ODE) \cite {Fraed79} with ice-albedo 
feedback taking into account the albedo of melt ponds. We modify this model assuming a
stochastic  distribution of melt pond sizes, based on the Fokker-Plank equation. 
After that we investigate equilibria of the resultant stochastic ODE under the key 
assumption that the surface temperature is a slow function of time relative to melt 
pond size. Different bifurcation regimes were obtained for this model. One of 
them may be quite interesting for climate applications, where the temperature of this 
system is stabilized only due to the fractal transition in melt pond geometry.

\section{Evolution of melt ponds}

\subsection{Mechanism of the fractal transition}

\begin{figure}[t]
\center{\includegraphics[width=1.0\linewidth]{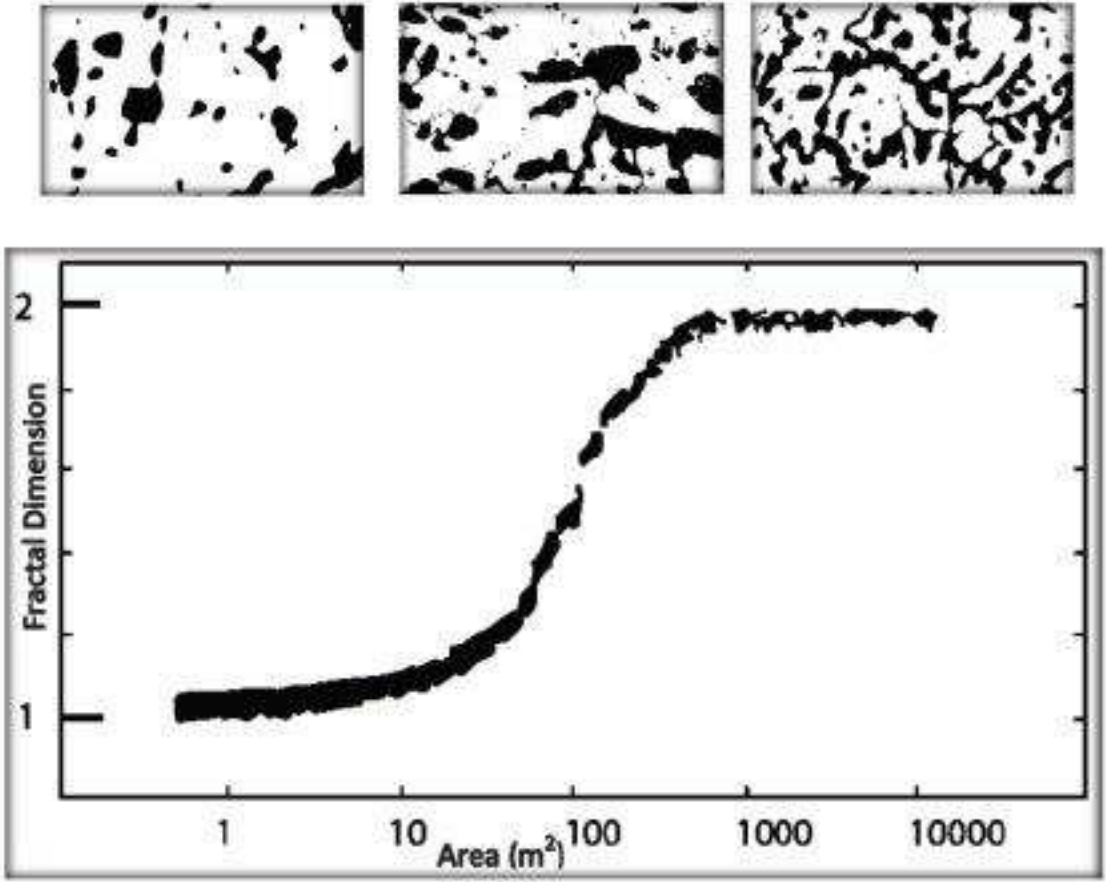}}
\caption{Melt pond fractal dimension $D$ as a function of area $S$, showing the transition to complex ponds with increasing length scale. Ponds corresponding to the three regimes are shown to upward: small ponds with smooth boundaries and $D\approx 1$, transitional ponds with a horizontal scale of about $50$ m, and complex ponds with river-like boundaries with $D\approx 2$. Adapted from \cite{Hoh12} with permission.
\label{Fig2}}
 \end{figure}

Viewed from high above, the sea ice surface can be thought of as a two phase composite of ice and melt water. The boundaries between the two phases evolve with increasing complexity and a rapid onset of large scale connectivity, or percolation of the melt phase (Fig.\ref{Fig2}).
As  was shown in \cite{Hoh12} that the  
melt pond perimeter $\Pi$ can be  defined approximately by
\begin{equation}
\label{Eq(23)}
\Pi\sim\sqrt{S}^{\;D},
\end{equation}
here ${S}$ is the area of  ponds and $D$ is the the fractal dimension.
The authors have observed a transition 
from $D = 1$ to $D\approx 2$ as the ponds grow in size, 
with the transitional regime centered around $\mathrm{{100 \; m^{2}}}$.
According to \cite{Hoh12} there exist  three regimes:

\vspace{1ex}

A)  $S<\mathrm{{10 \; m^{2}}}$; then we observe simple ponds with smooth boundaries and $D\approx 1$;

B)  $\mathrm{{10 \; m^{2}}} < S < \mathrm{{1000 \; m^{2}}}$; corresponding to  transitional ponds where complexity increases rapidly with size;

C)  $S > \mathrm{{1000 \; m^{2}}}$; complex, self-similar case, where pond boundaries behave like space filling curves with $D\approx 2$ (so-called fractals).

Here, we can 
show the transition in empirical formula (\ref{Eq(23)}) can be 
obtained from the rigorous pattern formation theory. To this end, we 
use the Kuramoto-Sivashinsky equation \cite{Lan87} that allows us to  
demonstrate that beginning with a critical characteristic size, the boundaries
become unstable with respect to perturbations along the boundary. 

To describe the beginning of the fractal boundary growth, we can use the linearized Kuramoto-Sivashinsky equation:
\begin{equation}
\label{Kur}
\tilde h_t= -m_0 \tilde h_{zz} -  n_0\tilde h_{zzzz},
\end{equation}
where $\tilde h$ is a normal displacement along the boundary; $z$ is the coordinate along the boundary; $z \in [0, P]$; $P$ is a pond perimeter, $m_0$ and $n_0$ are positive coefficients. Since the pond boundary is a closed curve,  
we set $P-$periodic boundary conditions for
$\tilde h$.
Then the nontrivial solution of Eq. (\ref{Kur}) is
\begin{equation}
\label{Eq(58)}
\tilde h=P_0\exp(ik z+ \beta_0(k) t),  \quad  \beta_0 =m_0 k^2 - n_0 k^4,
\end{equation}
where $k=2\pi m_1/P$, $m_1$ is a positive integer, and $P_0$ is a constant.
Hence, the minimum $k$ is $k_{min}=2\pi/P$.  If $\beta_0(k_{min}) < 0$  for all $k$ then
the boundary is stable because all perturbations decrease exponentially.
If $\beta_0(k_{min}) > 0$  we have an instability and $\tilde h$ increases with an exponential rate. 
Clearly, it must be a sufficiently large characteristic size. It is well known that the Kuramoto-Sivashinsky equation describes fractal growth \cite{Lan87}. 
We thus obtain the next equation that defines the critical perimeter found in \cite{Hoh12}:
\begin{equation}
\label{Eq(59)}
P_c=2\pi (n_0/m_0)^{1/2}.
\end{equation}

According to this assumption, we can suppose pond boundaries with fractal dimension about one can be considered like growing elliptical curves (there are circular ponds, in the ideal case) which become unstable at some characteristic size $R$, the length of the semi-major axis: $a_e=r_{e1}R$, $b_e=r_{e2}R$, where $a_e$ and $b_e$ have the same order. In the case of fractal transition, ponds are close to long and narrow ellipses, where $a_e$ and $b_e$ have the different order. These ellipses remind one of rivers rather than the simple circular ponds (Fig.\ref{Fig2}). Then one can expect that the area of such a river of length $R$ is proportional to $R$.
 
\subsection{Melting front of pond}

\begin{figure}[t]
\center{\includegraphics[width=1.0\linewidth]{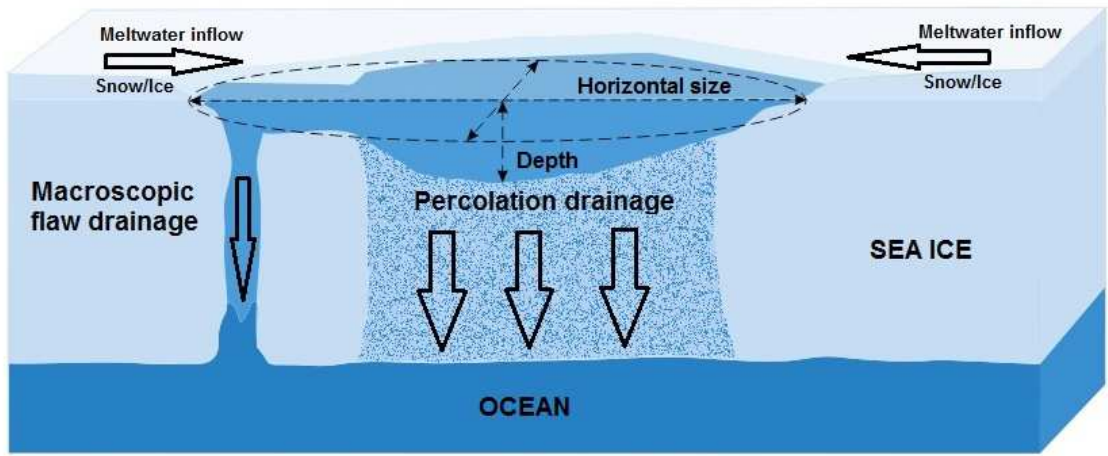}}
\caption{Schematic representation of a melt pond. Adapted from \cite{Pol12} with permission.
\label{Fig3}}
 \end{figure}

Our initial 
considerations of melt ponds will be based 
on the following geometrical property of melt ponds.
Typically, developed ponds have \cite{Fet98,Pol12} horizontal (characteristic) sizes ($R$) 
on the order 10--1000 m, and a small depth (z) of 0.1--0.8 m, 
i.e. a melting layer has a small but non-zero thickness (see Fig.\ref{Fig3}). 
Specific geometric features of melt ponds are determined through fundamental physical processes in sea ice. The complexity of the hydrology and thermodynamics of melt pond 
formation is the basis for sophisticated 
numerical models of melt pond evolution  
\cite{Sky07,Felt2010}. We do not discuss here the details of 
the thermodynamic processes in sea ice leading to the 
formation of the melt ponds. However, we can determine melting front (corresponding to the length of the semi-major axis of the elliptical ponds), following  by a phase transition model \cite{caginalp 89} where the melting layer has a small but non-zero thickness and a large horizontal dimension that agrees with our problem. Also, we can suppose that the ice-water interfaces are quasi one-dimensional, following \cite{MolVak88} we obtain the relation for the melting front velocity: 
\begin{equation}
  v^{\star}(x,y,z,t)=\delta(T)
\label{MCur}
\end{equation}
where $v^{\star}$ is the normal melting front velocity at the point $(x,y,z)$
and $\delta$ is a function of melting surface temperature $T$. The quantity $\delta$ can be expressed via microscopic parameters of the phase transition problem \cite{caginalp89,  MolVak88, Fife77}, however, it is simpler to find this
quantity by experimental data since $\delta$ determines the main contribution in the pond area increasing.

We are planning to consider the planar case. In this case, our fronts are curves. All fronts are closed curves, 
which  initially are not too different from ellipses. 
For elliptical fronts 
of size $R(t)$, Eq. (\ref{MCur}) takes the form
\begin{equation}
\frac{dR}{dt}=\delta (T), \label{MCur1}
\end{equation}

Some actual melt ponds can be thought of as three dimensional lenses (see Fig.\ref{Fig3}). In \cite{Pol12} some important effects are described and experimental data are presented.  
It is shown that there is a vertical transfer of water in the ocean through ice percolation, permeability, or macroscopic flows, which is proportional to the depth of the lens.
We can assume that on average this depth is proportional to the pond size $R$.  Therefore,
due to this effect, a rough estimate of the rate $dW/dt$ of the water mass in the pond  is $-\beta W$.
Since $W = const R^3$, we have the following contribution $R_{w}$ of this effect into
$dR/dt$:  $R_{w}=-\gamma  R$. Taking into account this effect, we change Eq.(\ref{MCur1}) into the form
\begin{equation}
\frac{dR}{dt}=\delta(T) -\gamma(T) R =P(R, T), \label{MCur2}
\end{equation}
where we suppose that $\delta$ and $\gamma$ depend on the temperature.

\subsection{Melt ponds and albedo}

Albedo is the reflecting power of a surface. 
Material properties, surface topography, and other properties of the surface 
influence albedo as well as related
feedback mechanisms. We will involve 
melt ponds in the feedback by means of area. 
For this aim we apply formula (\ref{Eq(23)})
to study  the melt ponds area. 

The total average albedo $A$ can be approximated by
\begin{equation}
\label{albedo}
A=A_{rp}\frac{S_{rp}}{S_{rp} + S_{arc}} +  A_{arc} \frac{S_{arc}}{S_{rp} + S_{arc}}, 
\end{equation}
where $S_{arc}$ is the   area of the Arctic zone covered by ice for low temperatures  and $S_{rp}$ is the area of the  rest planet, $A_{rp}$ is the average albedo of the rest planet, 
 and $A_{arc}$ is the average albedo of Arctic zone. 

According \cite{Fet98}, the albedo of the Arctic surface is
\begin{equation}
\label{Eq(26)}
A_{arc}=A_0(1-S_{r})+B_0 S_{r} = A_0-(A_0-B_0) S_{r},
\end{equation}
where  $A_0$ is an average albedo of ice area, $B_0$ is an average albedo of melt ponds, the percentage of the surface covered by ponds:
$S_r=\frac{S_{melt}}{S_{arc}}$ with ${S}_{melt}$ -- the average area of all melt ponds. 
Thus, we have obtained the formula for albedo involving the area of the surface covered by melt ponds. 



Using the facts about the fractal transition we compute the melt pond area as follows. For the
averaged size $R(t) <R_F$, again  we assume that shape of melts ponds are close to ellipses. Then
we define the area of melt ponds by
\begin{equation}
S_{melt}({\bf R}) \approx \pi c_1 \sum_{i=1}^N R_i(t)^2, \label{SI1}   
\;\;\;\;\; S_{melt} < s_*N \approx N \pi c_1 R_F^2
\end{equation}
the coefficient $c_1$ takes into account a deviation of elliptical form, $R_F$ is a critical characteristic size of melt pond at the fractal transition. Here
${\bf R}=(R_1(t), ..., ... R_N(t))$ is a vector of pond sizes
and $N$ is the number of the ponds.
For $R > R_F$  the ponds can be envisioned as long and narrow (albeit contorted). 
Then we use the
relation
\begin{equation}
S_{melt}({\bf R}) \approx c_2 \sum_{i=1}^N  R_i(t), \label{SI2}  
\;\;\;\;\;   S_{melt} > s_*N,
\end{equation}
where $c_2$ is a constant, which determines a  characteristic average width of river-like ponds (that we observe after
the fractal transition).  



\section{Low order climate model with ice-albedo feedback for melt ponds}

In the previous sections, we have obtained expressions 
for the albedo involving the percentage of the surface covered
by melt ponds, which depends on the area of the ponds. 
In turn, the area evolution depends on melt pond  dynamics. 
This  can be exploited in a conceptual climate model.  
Such models are based on an ice-albedo feedback that allows 
albedo to be temperature dependent. These models couple the 
albedo to the global energy balance through inclusion 
of heat transport \cite{Cur95,Nor75}.
In this section we show how these  
models can be developed taking into account melt pond characteristic size dynamics. 
It is  based on a relationship between albedo, 
melt pond size, and temperature. It allows us to 
find a climate bifurcation point related to melt ponds, 
and estimate climate sensitivity provided that melt ponds 
play a key role in the mechanism of ice-climate feedback.

A simple climate model is a one-dimensional system which can be described \cite{Fraed79} by 

\begin{equation}
\label{Eq(30)}
\frac{dT}{dt}=\frac{1}{\lambda}(-\epsilon \sigma T^{4}+\frac{\mu_0  I_0}{4} (1-A)),
\end{equation}
where $\lambda$ is thermal inertia, $T$ is surface temperature, $t$ is time, and $A$ is the albedo of the surface. The left term characterizes the time-dependent behavior of the climate system, usually taken to mean an average surface temperature. Surface temperature changes as a result of an imbalance in radiative heat transfer. On the right hand side, the first term is outgoing emission and the second term represents incoming solar radiation. Generally, incoming solar radiation to earth's surface should depend on total solar radiation incident on earth ($\mu_0$), and the solar constant $(I_{0})$ as well as surface albedo. On the other side, outgoing emission can be described through the fourth power of temperature, the effective emissivity ($\epsilon$) and a Stefan-Boltzmann constant ($\sigma$).

Substituting the formula (\ref{albedo}) via the pond characteristic size, we finally have the following system for $R_i, T$:
\begin{equation}
\label{Eq(31)}
\frac{dT}{dt}=f({\bf R}, T),
\end{equation}
where the right hand side is a sum of two terms that describe,  the contributions of land albedo, land emissivity and arctic albedo, respectively: 
$$
f({\bf R}, T)=F_{rp}(T) +  F_{arc}(S_{melt}({\bf R})),
$$
where
\begin{equation}  \label{LandTerm}
F_{rp}(T) =\frac{1}{\lambda}(-\epsilon \sigma T^{4}+
\frac{\mu_0}{4} I_{0}(1-A_{rp}(T)),
\end{equation}
 \begin{equation}  \label{ArcTerm}
F_{arc}({\bf R}, T)=\frac{\mu_0 I_0}{4 \lambda } (A_0- (A_0-B_0)S_{melt}({\bf R})/S_{arc}).
\end{equation}
Here we assume, for simplicity, that $A_{rp}(T)$ is a regular function of averaged temperature $T$, which weakly depends on $T$ at some value $T_s$. This value defines the averaged surface temperature for 
the case when all Arctic is covered by ice,  i.e., $S_{melt}=0$:
\begin{equation}
\label{T0}
F_{rp}(T) +  F_{arc}(0)=0
\end{equation}
 (following here ideas from \cite{Resonance} and \cite{Berglund}). 

For $R_i$ we use the equation
\begin{equation}
dR_i=P(R_i, T)dt + 2\kappa d\omega_i, \label{Lang}
\end{equation}
where $i=1,..., N$ and $\kappa$ is a parameter. 
Here, observing
that pond growth can be viewed as a stochastic process as was presented in \cite{Yack00}, 
we use the Langevin equation for $R_i$, 
where $d\omega_i$ are independent standard Wiener
processes. Since $N >> 1$ we can also use  
the Fokker-Planck equation for the pond size distribution $\rho(R_i,t)$:
\begin{equation}
\frac{\partial \rho}{\partial t }=-\frac{\partial P(R_i, T) 
\rho}{\partial R_i} + \kappa^2 \frac{\partial^2 \rho}{\partial {{R_{i}}^{2}} }.  
\label{FP}
\end{equation}

This model involves the additive noise  generated by the term $\kappa d\omega_i$. We need such a term in order to obtain a reasonable pattern of pond sizes for large $t$ since
otherwise we obtain that all the ponds have  the same size as $t >>1$. Moreover,  the stochastic model allows us to describe stochastic resonance effects \cite{Resonance, Berglund}, which are possible
here.  

This nonlinear climate model can be reformulated as a 
stochastic dynamical system. Note that
for $R_i > R_0$, $\kappa=0$ (when stochastic effects are absent), 
and with increasing $\delta(T)$ one has
$$
\frac{\partial f(R_i,T)}{\partial R_i} > 0,  \quad \frac{\partial P(R_i,T)}{\partial T} > 0.
$$
This means that the system (\ref{Eq(31)}) 
is cooperative. Therefore,
due to fundamental results of M. Hirsch \cite{Hir84}, 
this system cannot exhibit oscillating  
solutions and the Andronov--Hopf bifurcations
\cite{Arn83}.
All trajectories converge to equilibria and
the attractor is a union of these equilibria.

This observation allows us to compute the pond area $S_{melt}$ for large times.  
In physically realistic situations $N >> 1$, 
so we can simplify the approximations Eqs.
(\ref{SI1}) and (\ref{SI2}).  We can transform these relations as follows
\begin{equation}
S_{melt} \approx S_c=\pi c_1 N \int_0^{\infty} {{R_{i}}^{2}} \rho(R_i, t)dR_i, \label{SI1C}
\end{equation}
for $S_c < s_* N$, where
$s_*=\pi c_1 R_F^2$, and $c_1$ is a constant taking into account 
the deviation from the elliptical pond form.

After the fractal transition
one has
\begin{equation}
S_{melt} \approx S_F=c_2 N \int_0^{\infty}  R_i \rho(R_i,t)dR_i, \label{SI2C}
\end{equation}
for $S_c > s_* N$. Here $\rho(R_i, t)$ can be  defined by Eq. (\ref{FP}).

\section{Analysis of the system for temperature and ponds}

Equilibria of the system (\ref{Eq(31)}) and (\ref{Lang}) 
for $\kappa=0$ can be found as follows. 
For fixed temperature $T$ we compute
quasi-equilibria $R_i(T)$  setting $P(R_i, T)=0$. 
This equation  has the root
\begin{equation}
  R_+(T)=\frac{\delta(T)} 
{\gamma} \label{Rstate1}
\end{equation}

Note that the root $R_+$ is a
stable resting point (a local attractor) of a semi-flow 
defined by Eq. (\ref{Eq(31)}). Therefore, the dynamics 
of Eq. (\ref{Eq(31)}) can be described as follows:   $R_i(t) \to R_{+}$ for large 
times.

{\em Our key assumption} is  that $T$ is a slow function of time 
relative to $R_i(t)$, i.e., the melting 
process for ponds is fast while
changing of the related climate system is slow.

Under this  
assumption computing equilibria for the
temperature $T$ becomes a mathematically tractable problem even in 
the stochastic case $\kappa >0$.
In fact, then (using classical results of dynamical 
systems theory) we  solve the Fokker-Planck equation (\ref{FP}) 
for each fixed $T$,  after which we substitute the
results in Eq. (\ref{Eq(30)}) and find the equilibria for $T$.  
So, let us fix $T$ in Eq. (\ref{FP}). It is well know that
$\rho(R_i, t) \to \rho_{eq}$ for large times $t$, where 
$\rho_{eq}$ is an equilibrium distribution defined by
\begin{equation}
\rho_{eq}= C(T) \exp(- \kappa^{-2} V(R)), \quad
\end{equation}
with
$$
V(R)= \delta(T)R  -0.5 \gamma(T) R^2,
$$
 where $C(T)$ is a factor such that 
$\int_0^{\infty} \rho_{eq}(R) dR=1$.
 We have then
  \begin{equation}
S_{melt}=\pi c_1 N C(T)\int_0^{\infty}  
R^2 \exp(- \kappa^{-2} V(R))dR, \label{SI1E}
\end{equation}
before the fractal transition   and
\begin{equation}
 S_{melt}=c_2 N C(T) \int_0^{\infty}  
R \exp(- \kappa^{-2} V(R)) dR, \label{SI2E}
\end{equation}
after this transition.
Therefore, for small $\kappa$  we obtain 
the following relations for the pond area 
$S_{melt}$ (using that the function $\rho_{eq}$ 
is well localized at $R =R_{+}(T)$)
\begin{equation}
S_{melt}(T)=C_0 N   (R_{+}(T)) ^2, \label{SEDF}
\end{equation}
for $R_{+}(T) < R_F$,  and
\begin{equation}
 S_{melt}(T)=C_0 N R_{+}(T) R_F \label{SEF}
\end{equation}
for $R_{+}(T)  \ge  R_F$.  Here $C_0$ is a constant 
and $R_F$ is a critical characteristic size of melt pond at the fractal transition.
We assume that $R_+(T)$ is an increasing function of  $T$, i.e., 
$
 \frac{dR_{+} (T)}{dt} >0.
$
This assumption looks natural.
Note that $S_{melt}(T)$ has such properties.  
This function is continuous and has a derivative $dS_{melt}/dT=S_{melt}'(T)$,
which has a break at the temperature $T_F$
such that $S_{melt}(T_F)=R_F$. 

For the temperature $T$, as a result of some straight forward transformations, we obtain then the evolution equation
\begin{equation}
\label{Tdiff}
\frac{dT}{dt}=G(T),
\end{equation}
where
\begin{equation}
\label{gT}
 G(T)=  \zeta(T)    - Q(T),
\end{equation}
There are
$$
\zeta(T)=\frac{4 \epsilon \sigma T^4  (S_{rp} + S_{arc})}{\mu_0 I_0 S_{arc}} +  A_{rp}(T)\frac{ S_{rp}}{S_{arc}} +  A_0
$$
and
$$Q(T)=(A_0 - B_0) \frac{{S_{melt}(T)}}{S_{arc}}.$$

The equilibria of this equation are defined by
\begin{equation}
\label{Teq}
  Q(T)=\zeta(T).
\end{equation}
These equilibria are intersections $T_{eq}$ of the curves 
$\zeta(T)$ with $Q(T)$.
If for $T_{eq}$ one has
$$
\zeta'(T_{eq}) < Q'(T_{eq}),
$$
then the intersection gives us a stable equilibrium and 
and thus a local attractor, otherwise
this equilibrium is a saddle point.  

Note that the 
function $Q(T)$ equals zero for $T < T_b$, where $T_b$ is the temperature of the phase transition, when we have no melt ponds, it grows faster in $T$ for smaller $T$ 
while the averaged pond size is less than the critical value around
$R_F$. This means that early in the warming cycle we 
observe fast growth and afterwards when the ponds become 
fractals, the growth of $Q(T)$ in $T$ is slower.
This result is consistent with experimental data 
\cite{Pol12}.  

\begin{figure}[t]
   \centering
\includegraphics[width=1.0\linewidth]{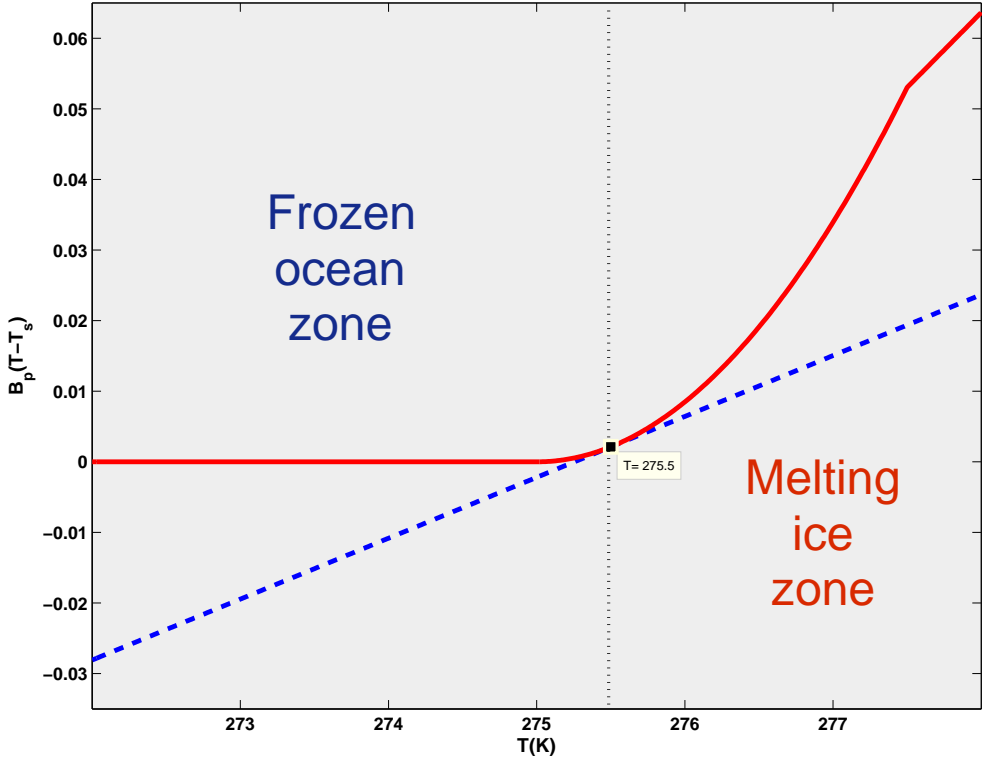}
\caption{This picture illustrates the bifurcation, here  $T_{eq}=275.50$ K is a single equilibrium value. The dotted blue line corresponds to the term  $B_p(T-T_s)$ in the equation (\ref{quadr}) and the red curve is $q(T)=(A_0-B_0)\frac{S_{melt}(T)}{S_{arc}}$. The steady state value of $T$ can be obtained as  intersections of these curves.}
\label{Fig6}
\end{figure}

The analysis of Eq.(\ref{Teq}) can proceed if we take into account that $S_{arc} << S_{rp}$ and, moreover, supposing that melting phenomena appear at some temperature interval $T_0,  T_1$, following \cite{Resonance, Berglund} note that 
$T^4$ and $A_{rp}(T)$ vary insignificantly on this range. The  the problem can be further simplified by
a linearization of
 $\zeta(T)$  at the temperature $T_s$ which is an equilibrium averaged surface temperature of the system ``The rest of the planet + the Arctic  zone'', where $S_{melt}=0$.  We have
$$    
\zeta(T_s)=0
$$
and, following \cite{Abbot}, consider $Q(T)$ as a small but sufficiently irregular in $T$ perturbation. By an elementary perturbation theory, we have  that the temperature $T$ is defined by
\begin{equation} \label{TeqP}
\zeta'(T_s) (T- T_s)= Q( T).
\end{equation}
Depending on the parameters $A_0- B_0$,  $S_{arc}/S_{rp}$,     
$\beta=\frac{ \mu_0 I_0}{4 \epsilon \sigma }$, 
 and others there are possible such main cases:

({\bf I}) a single stable equilibrium which serves as a global attractor, Fig. \ref{Fig6};

\begin{figure}[t]
   \centering
\includegraphics[width=1.0\linewidth]{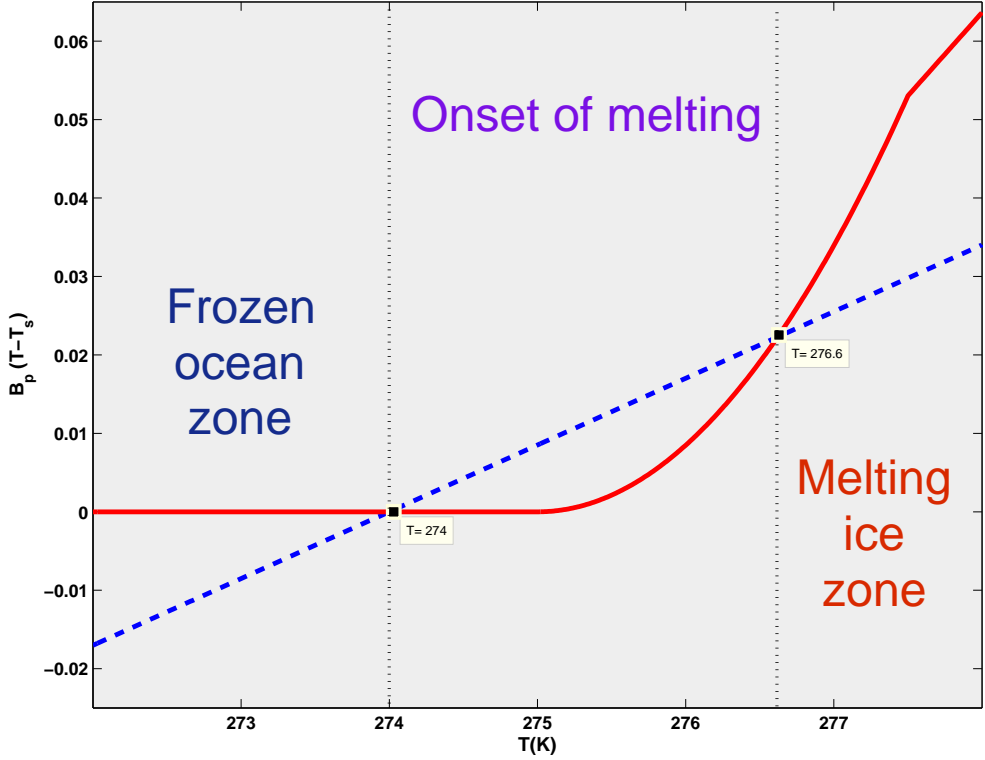}
\caption{Here we see  the case of two equilibria.  We have a stable equilibria at $T_{eq}=274.00$ K  and a unstable one at  $T_{eq}=276.60$ K 
(for  stable equilbria $T_{eq}$ one has  $B_p > q^{'}(T_{eq})$). The dotted blue line corresponds to the term  $B_p(T-T_s)$ in the equation (\ref{quadr}) and the red curve is $q(T)=(A_0-B_0)\frac{S_{melt}(T)}{S_{arc}}$. The steady state value of $T$ can be obtained as  intersections of these curves.}
\label{Fig4}
\end{figure}

({\bf II}) a stable and unstable equilibria, Fig. \ref{Fig4}; 

\begin{figure}[t]
   \centering
\includegraphics[width=1.0\linewidth]{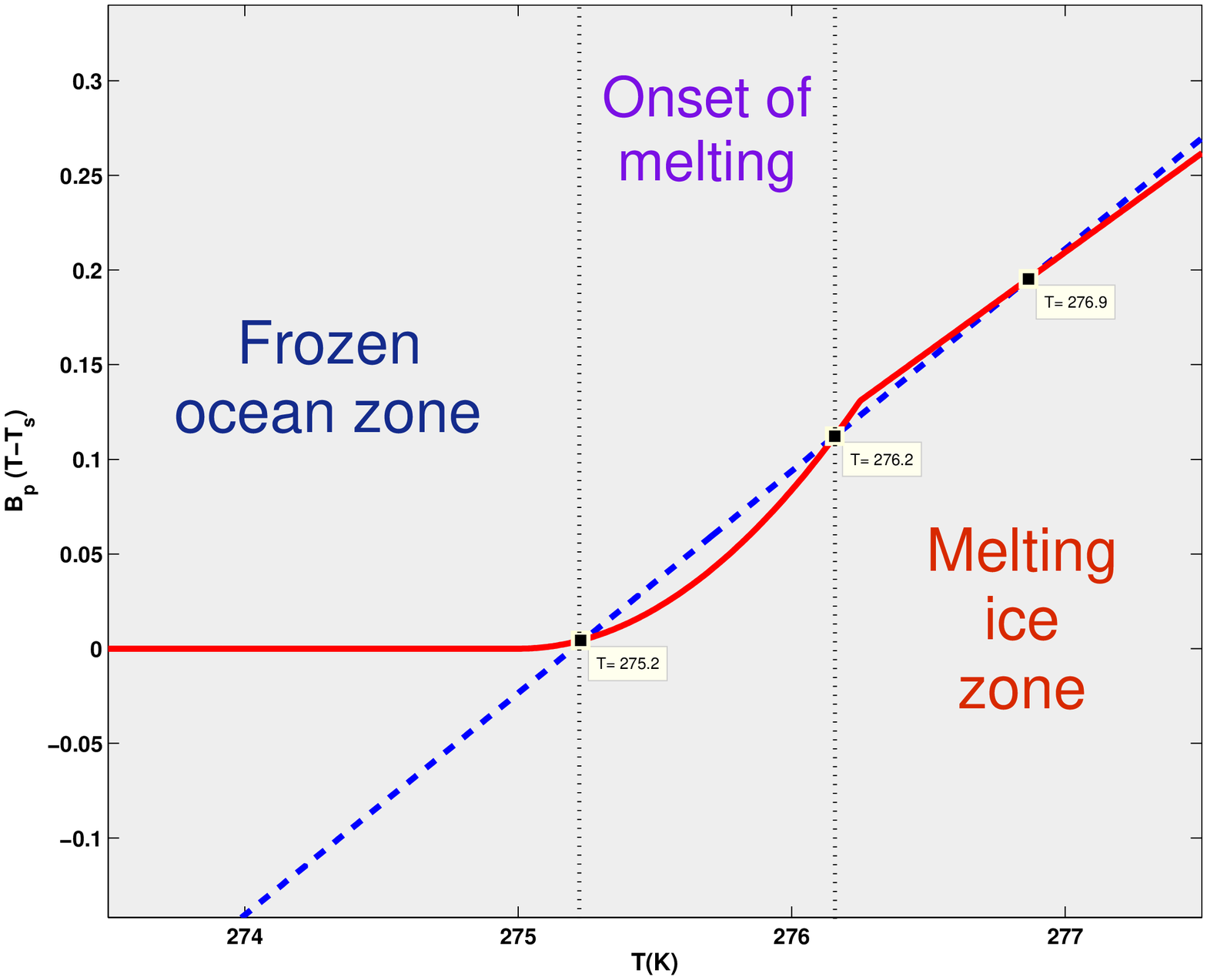}
\caption{This plot shows the case of three equilibria. A stable equilibria $T_{eq}$  are  $275.24$,  $276.20$ and  $276.90$ K. The blue dotted line corresponds to the term  $B_p(T-T_s)$ in the equation (\ref{quadr}) and the red curve is $q(T)=(A_0-B_0)\frac{S_{melt}(T)}{S_{arc}}$. The steady state value of $T$ can be obtained as  intersections of these curves.}
\label{Fig5}
\end{figure}

({\bf III}) two stable equilibria plus a saddle point, Fig. \ref{Fig5}.

A  bifurcation picture occurs if we  
assume that $Q(T)$ is close to $0$  for $T < T_{b}$,
and increasing for $T > T_b$. This condition looks natural since
for low temperatures melt ponds are frozen.

We can take the following approximation, when Eq. (\ref{TeqP}) can be solved analytically.  Let us set
$S_{melt}=0$ for $T < T_b$. For $T> T_b$ we use relations
  (\ref{SEDF}) and (\ref{SEF}) with some $C_0 \approx\pi$ and 
$R_+(T) \approx r_0(T- T_b)$, where $r_0 >0$ is a parameter, which determines the pond size increase in temperature $T$. 
 Such an approximation means that
we use linear approximations for $\delta(T)$ and $\gamma=const$ for $T > T_b$. Then  Eq. (\ref{TeqP}) becomes
\begin{equation} \label{quadr}
\begin{split}
          B_p(T-T_s)=(A_0-B_0)\frac{S_{melt}}{S_{arc}},
\\ \quad  B_p=\frac{4\epsilon\sigma T_s^3} {I_0 \mu_0/4} - a_p, \quad a_p=\frac{dA_{rp}(T)}{dT}\vert_{T=T_s},
\end{split}
\end{equation}
where the right hand side is zero for $ T < T_b$, it is quadratic function in $T$ for $ T \in (T_b, T_F)$ and it is a linear function for $T > T_F$.


Note that Eq.(\ref{quadr}) can have $n=1$, $n=1, 2$ or $n=3$ roots and it is also possible that roots are absent.  
If the fractal transition is absent (or $R_F$ is too large),  then either there are no roots or $n=2$, in the second case only a single root is stable.     
When the fractal transition exists, we can have $n=3$ if $B_p$ is larger some critical level  $ B_p^*=C_0 N R_F r_0/S_{arc}$. One node $n=1$ is possible too, in this case we have one stable point when the system is moving from one state to another state that is not due to the fractal transition. 
 We conclude that the transition from
two solution to three occurs if the
parameter $B_p$ changes but only when the fractal transition exists. In this case, a pitchfork bifurcation is possible. If the fractal transition
does not occur,  a saddle-node bifurcation can appear, when $n=0$ or $n=2$ and we have a single equilibrium ($n=1$) at the bifurcation point.

For  $\tau=T_s-T_b>0$  three equilibria  are possible if and only if the following conditions hold:
$$
v \in (1/2,1),  \quad  u >  v(1-v), \quad   u < 1/4,   \quad       
$$
where
$$
b=(A_0 -B_0)C_0 N  r_0^2/(B_p  S_{arc}),    \quad u=b\tau,   \quad   v=b(R_F/r_0). 
$$

Here, we list the parameters, which were used for Figs. \ref{Fig6}, \ref{Fig4}, \ref{Fig5}. There  are effective emissivity $\epsilon=0.62$, average albedo of ice area $A_0=0.68$, Stefan-Boltzmann constant $\mathrm{\sigma=5.67\cdot 10^{-8} \; J\cdot s^{-1} m^{-2} K^{-4}}$
average albedo of melt ponds $B_0\approx0$, $\mu_0=1.00$  and incoming solar energy $ I_{0}/4$ is $\mathrm{340.00 \; W\cdot m^{-2}}$. We have put $\mathrm{S_{arc}=5.00\cdot 10^{12}\; m^2}$ and $a_p=0$. The number $N$ of the ponds is $N\approx 4.00 \cdot 10^8$. In case of Fig.\ref{Fig6} there are $R_F=35.00$ m, $r_0=3.00$ m/K, $T_b=275.00$ K, $T_s=274.50$ K. For Fig. \ref{Fig4} the parameters are $r_0=7.00$ m/K, $T_s=276.00$ K, $T_b=275.00$ K and $R_F=17.50$ m. The parameters $R_F=25.00$ m, $r_0=20.00$ m/K, $T_b=275.00$ K, $T_s=275.20$ K are used in Fig. \ref{Fig5}.

\section{Discussion}

In this section, we discuss some physical consequences of the obtained results. 

\subsection{Melt pond evolution and sea ice melt pond area.}
First of all, we discuss the different regimes of our toy climate system related to bifurcations that can happen in this system.
In the case of saddle-node bifurcation we have two stable zones: ``Frozen ocean'' and ``Melting ice'', see Fig.\ref{Fig6}. Such kinds of climate states were described in the earliest works \cite{Cur95,Nor75}. We can explain its existence through the phase transition. However, the other two cases are more interesting.

In the case of two equilibria (Fig.\ref{Fig4}), we can distinguish three different zones. Two of them are similar to the first case ``Frozen ocean'' and ``Melting ice'', however we introduce a new zone between two equilibria: ``Onset of melting''. This zone corresponds to the initial growth of melt ponds with the  elliptical shapes. Physically, the existence of this zone plays an important role, because seasonal sea ice minimum strongly correlates with beginning melt pond fraction as was shown in \cite{Sch14}, based on statistical analysis of data from models. In this paper, it is  shown that this zone, which is located between two stable and unstable equilibria,  determines the future state of this system. However, here we can suppose that in the ``Melting ice'' zone growing ponds will cover a significant ice surface that will lead to full ice disappear in during one season. 

In the case of three equilibria (Fig.\ref{Fig5}), we still distinguish three different zones: ``Frozen ocean'' subsists due to the low temperature; ``Onset of melting'' still exists, however in this case it is shorter, because the  elliptical ponds shifts its shapes very fast to narrow and long rivers due to the fractal transition, which corresponds to the second point of equilibrium. Crossing this point, the melt ponds are approaching to the complex fractal forms. Such fractal system  stabilizes our simple climate system at the third point of equilibrium. After that point the pond growth is absent. Computations show that in the second case the area {$S_{melt}$ covered by ponds is $\mathrm{{1.60\cdot10^{11} \; m^{2}}}$ at $T_{eq}=276.60$ K, but in the third case the area {$S_{melt}$ covered by ponds is significantly less $\mathrm{{0.27\cdot10^{11} \; m^{2}}}$ at $T_{eq}=276.90$ K. Thus, we can conclude melt ponds help to prevent full summer Arctic sea loss, because they can stabilize the state of the climate system due to the fractal transition. In addition, existence of ``Onset of melting'' zone due to the transition from stable to unstable equilibria allows to control the amount of sea ice extent by the end of the melting ice season. 

In addition, we can consider another parameter which can control the physical state of the system: there is a thermal inertia $\lambda$ (Eq.\ref{Eq(30)}). A huge heat capacity of the ocean produces the thermal inertia that can make surface of melting or freezing more gradual. In our model the parameter $\lambda$ defines a rate of the system approaching to an equilibrium. Usually, conceptual models take into account this parameter as a constant, however in case of melt pond incorporated models this parameter may be defined as a function, then the rate of reaching equilibrium will be easily computed. It can help to understand how fast a bifurcation may happen. However, these models should incorporate more complicated thermodynamics of the ocean-atmosphere interaction, at least, a model such as was suggested in {\cite{Eis09}.

\subsection{Melt ponds evolution and critical behavior of albedo.}

From the first appearance of visible pools of water, 
often in early June, the area fraction of sea ice 
covered by melt ponds can increase rapidly to over 70 percent
in just a few days. Moreover, 
the accumulation of water at the surface dramatically 
lowers the albedo where the ponds form. A corresponding 
critical drop-off in average albedo \cite{Pol12}. The resulting increase in solar absorption 
in the ice and upper ocean accelerates melting \cite{Per03}, 
possibly triggering ice-albedo feedback. Similarly, 
an increase in open water fraction lowers 
albedo, thus increasing solar absorption and 
subsequent melting. The spatial coverage and 
distribution of melt ponds on the surface of ice 
floes and the open water between the floes thus 
exerts primary control of ice albedo and  
partitioning of solar energy in the ice-ocean 
system \cite{Eic04,Pol12}. 

Thus, each data set exhibits critical behavior at the onset of melt pond formation,
similar to the behavior of order parameters characterizing phase transitions in thermodynamics.


We would like to discuss such critical behavior related to melt pond evolution based on our model. Here, we are taking into account that the melt pond size is a fast variable, and the surface temperature (time-averaged) is a slow variable. Therefore, the mean size depends on the temperature. Also, $\delta(T)$ and $\gamma (T)$ in Eq. (\ref{MCur2}) are close to constant (or slightly changing as a function of $T$). In this case, the size is a smooth function of the temperature. When the critical size is changing due to the fractal transition, functions of melt pond area 
have a jump at this point(see Eqs. (\ref{SI1}), (\ref{SI2})). 
According to formula (\ref{Eq(26)}) the albedo  
depends linearly on area. 
So, we can approximate albedo as a hyperbolic tangent (see Fig. \ref{Fig7}) of 
the average surface temperature($\bar{T}$):  
\begin{equation}
\label{Eq(61)}
A(\bar{T})\approx A_F+A_m\tanh(T_\Delta) 
\end{equation}
where $A_F$ is the albedo of the surface after the fractal transition, 
and $A_m$ -- the constant corresponds to the change in albedo due to the 
fractal transition, $T_\Delta$ -- changing in the surface temperature due 
to the fractal transition  \cite{Pol12}. Previously, this formula was 
introduced empirically, based on the observation data. However, we may 
see physical interpretation of this phenomenon: the transition in fractal dimension of melt ponds affects  the shape of the albedo curve.  
\begin{figure}[t]
   \centering
\includegraphics[width=1.0\linewidth]{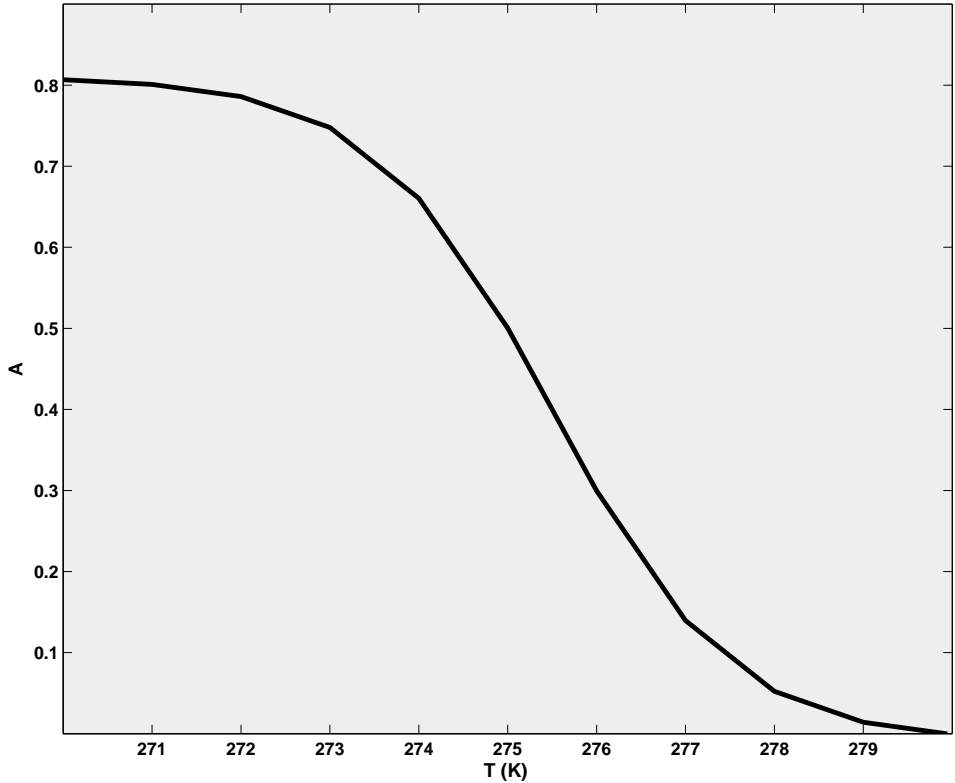}
\caption{Albedo as a hyperbolic function of the average surface temperature due to the fractal transition in the melt pond geometry.}
\label{Fig7}
\end{figure}

\section{Conclusions}

In this work, we have addressed  some fundamental questions related 
to the role of sea ice in  the climate system. 
First of all, we considered how 
geometrical properties of melt ponds 
can influence ice-albedo feedback and how 
it can influence  the bifurcation structure
of a simple climate model. 
The melting pond growth  model is developed to study melt pond 
formation and its changes in geometry. The approach,  proposed here,  can be 
useful for futur investigations of the geometry of melt 
ponds and their evolution.

We reviewed a low-order energy 
balance climate model using standard methods of dynamical 
systems theory. As a result, we  see different behavior of the climate 
system in the case of the ice-albedo feedback with 
melt pond following a stochastic distribution for the sizes.    
We concluded that in this case melt ponds  can strengthen the 
positive feedback and lead the climate system through  a
bifurcation point.   Moreover,
the melt pond contributions  can have a significant  
influence on the temperature state of the climate system.

We would like to emphasize that in this research three scales 
of the problem were connected. We have tied up micro, macro and 
global scales through the relation for albedo. Albedo 
(global scale) contains the area of melt ponds (expressed 
through sizes -- macro scale) which in turn is connected 
to the microscopic parameters describing thermodynamic changes in the melting front. Thus, this research  advances the multiscale  
approach to tipping point investigations, first presented for 
permafrost lakes in  \cite{Sud14}. 


\section*{Acknowledgments}
We gratefully acknowledge support from the Division of Mathematical
Sciences and the Division of Polar Programs at the U.S. 
National Science Foundation (NSF) through Grants
DMS-1009704, ARC-0934721, and DMS-0940249. We are also grateful for 
support from the Office of Naval Research (ONR) through
Grant N00014-13-10291. We would like to 
thank the NSF Math Climate Research Network (MCRN) as well for their support
of this work. Finally, this research was also supported 
by the Government of the Russian Federation through mega-grant 074-U01,
President's grant MK-128.2014.1, and RFBR's grant 14-01-31053.

\end{document}